\RequirePackage{ifpdf}
\ifpdf 
\documentclass[pdftex]{sigma}
\else
\documentclass{sigma}
\fi

\def\sdf#1#2#3{#1_{#2}(t,{#3})}             
\def\sdf-#1#2#3{#1_{#2}(-t,{#3})}           
\def\qwe#1#2#3{{#1}_{#2},\dots,{#1}_{#3}}   

\def\bs#1#2{{#1}\setminus{#2}}             

\def\E^#1{{\buildrel #1 \over\vee}}

\newcommand{\RR}{\mathbb R}

\newcommand{\gA}{\mathfrak A}

\newcommand{\abs}[1]{\left\vert#1\right\vert}

\newcommand{\norm}[1]{\left\Vert#1\right\Vert}

\begin{document}

\allowdisplaybreaks
\renewcommand{\PaperNumber}{053}

\FirstPageHeading

\ShortArticleName{On Regularized Solution for BBGKY
  Hierarchy of One-Dimensional Inf\/inite System}

\ArticleName{On Regularized Solution for BBGKY
  Hierarchy\\ of One-Dimensional Inf\/inite System}

\Author{Tatiana~V. RYABUKHA}

\AuthorNameForHeading{T.V. Ryabukha}

\Address{Institute of Mathematics of NAS of Ukraine, 3
         Tereshchenkivs'ka Str., Kyiv-4,  01601 Ukraine}
\Email{\href{mailto:vyrtum@imath.kiev.ua}{vyrtum@imath.kiev.ua}}

\ArticleDates{Received October 31, 2005, in f\/inal form April 26,
2006; Published online May 14, 2006}

\Abstract{We construct a regularized cumulant (semi-invariant)
representation
    of a solution of the initial value problem for the BBGKY hierarchy
    for a one-dimensional inf\/inite system of hard spheres interacting via
    a short-range potential. An existence theorem is proved
    for the initial data from the space of sequences of bounded functions.}

\Keywords{BBGKY hierarchy; cumulant; regularized solution}

\Classification{82C05; 82C40}

\section{Introduction}
Last ten years new functional-analytical methods of investigation
of the initial value problem to the BBGKY hierarchy for inf\/inite
systems have been developed in the books
\cite{PGM02,CGP97,CIP94,Sp91}. As is well known when constructing
a solution of the BBGKY hierarchy of such systems, considerable
analytical dif\/f\/iculties arise \cite{Pe79,PG83,GR03,GRS04}.
Until recently, existence theorems have been proved for
one-dimensional systems in the case of a certain class of
short-range interaction potentials and three-dimensional case only
for hard spheres systems \cite{PGM02}.

While constructing the solution of the initial value problem for
the BBGKY hierarchy of the classical systems of particles with the
initial data from the space of sequences of bounded functions, one
is faced with certain dif\/f\/iculties related to divergence of
integrals with respect to conf\/iguration variables in each term
of an expansion of the solution \cite{Pe79} (see also
\cite{PGM02,CGP97}). The same problem arises also in the case of
the cumulant representation of the solution stated in \cite{GR02}.

In this paper, we propose a regularization method for the solution
of the BBGKY hierarchy  in the cumulant representation. Due to
this method, the structure of the solution expansions guarantees
the mutual compensation of the divergent integrals in every term
of the series. We establish convergence conditions for the series
of the solution
 and prove an existence theorem  of a  local in time
weak solution of the BBGKY hierarchy for the initial data from the
space of sequences of functions which are bounded with respect to
the conf\/iguration variables and exponentially decreasing with
respect to the momentum variables.

\section{Initial value problem for BBGKY hierarchy}
Let us consider a one-dimensional system of identical particles
(intervals with length $\sigma$  and unit mass $m=1$) interacting
as hard spheres via  a  short range pair potential $\Phi.$ Every
particle $i$ is  characterized by phase coordinates
$(q_{i},p_{i})\equiv x_{i}\in\mathbb{R}\times{\mathbb R}$, $i\geq
1.$ For the conf\/igurations $q_i\in\mathbb{R}^{1}$ of such a
system ($q_i$ is the position of the center of the  $i$th
particle), the following inequalities must be satisf\/ied:
$|q_i-q_j|\geq\sigma,$ $i\neq j\geq1$. The set $ W_n \equiv \{
\{q_1,\ldots,q_n\} \mathrel| \exists(i,j),\, i\neq
j\in\{1,\ldots,n\}: |q_i-q_j|<\sigma \} $ def\/ines the set of
forbidden conf\/igurations in the phase space of a~system of $n$
particles. The phase trajectories of such hard sphere system are
determined almost everywhere in the phase space $\{\qwe
x1n\}\in\mathbb{R}^n\times(\mathbb{R}^n\setminus W_n)$, namely,
outside a certain set~${\mathcal{M}}_{n}^0$ of the Lebesgue
measure zero~\cite{PGM02}. The initial data $\{\qwe
x1n\}\in\mathbb{R}^n\times(\mathbb{R}^n\setminus W_n)$ belong to
the set~${\mathcal{M}}_{n}^0$ if (a)~there is more than one pair
collision at the same moment of time $t\in(-\infty,+\infty)$ or
(b)~inf\/initely many collisions occur within a f\/inite time
interval.

We assume that the interaction between the hard spheres is given
by a
 potential $\Phi$
with a~f\/inite range $R$ such that the  following conditions are
satisf\/ied:
\begin{alignat}{3}
        &    (a) \ \ &&    \Phi\in C^2([\sigma,R]), \quad 0<\sigma<R<\infty,&\nonumber\\
        &    (b)& &   \Phi(|q|)=\left\{\begin{array}{ll}
                                    +\infty, \ &|q|\in[0,\sigma),\\[1mm]
                                    0,             &|q|\in(R,\infty),
                                    \end{array}
                             \right. &\nonumber\\
        &    (c)&&    \Phi'(\sigma+0)=0. &
\label{Phi}
\end{alignat}
We note that  conditions (\ref{Phi}) imply the estimate
\begin{gather}
\label{Phii}
    \abs{\sum_{i<j=1}^{n}\Phi(q_i-q_j)}\leq bn,
    \qquad
    b\equiv
    \underset{q\in[\sigma,R]}{\sup}
    \big|\Phi(q)\big|
    \big(\big[\tfrac{R}{\sigma}\big]\big),
\end{gather}
where $\big[\frac{R}{\sigma}\big]$ is the integer part of the
number $\frac{R}{\sigma}.$

The evolution of states of the  system under consideration
 is described by
the initial value problem for the BBGKY
hierarchy~\cite{PGM02,CGP97}
\begin{gather}
         \frac{\partial}{\partial t}
         F_s(t,x_1,\ldots,x_s)
           =\{ H_s, F_s(t,x_1,\ldots,x_s)\} \nonumber\\
           \phantom{\frac{\partial}{\partial t} F_s(t,x_1,\ldots,x_s) =}{}        +\int dx_{s+1}
            \left\{
             \sum_{i=1}^s
             \Phi(q_i-q_{s+1}),
             F_{s+1}(t,\qwe x1{s+1})
             \right\}\nonumber\\
\phantom{\frac{\partial}{\partial t} F_s(t,x_1,\ldots,x_s) =}{}
+\sum\limits_{i=1}^s
             \int\limits_{0}^{\infty}
             dP\,
             P
             \cdot
             \Big(
             F_{s+1}(t,x_1,\ldots,x_s,q_i+\sigma,p_i-P)
\nonumber\\
\phantom{\frac{\partial}{\partial t} F_s(t,x_1,\ldots,x_s) =}{}
           -F_{s+1}(t,x_1,\ldots, q_i,p_i-P,\ldots,x_s,q_i-\sigma,p_{i})\nonumber\\
\phantom{\frac{\partial}{\partial t} F_s(t,x_1,\ldots,x_s) =}{}
           +F_{s+1}(t,x_1,\ldots,x_s,q_i-\sigma,p_i+P)
\nonumber\\
\phantom{\frac{\partial}{\partial t} F_s(t,x_1,\ldots,x_s) =}{}
           -F_{s+1}(t,x_1,\ldots, q_i,p_i+P,\ldots,x_s,q_i+\sigma,p_{i})
            \Big),\label{BBGKY}
\\
\label{BBGKYt0}
         F_s(t,x_1,\ldots,x_s)
         \big|_{t=0}
       =F_s(0,x_1,\ldots,x_s),
         \qquad s\geq1,
\end{gather}
where $\{\cdot,\cdot\}$ is the Poisson bracket, $H_s$ is the
Hamiltonian of the $s$ particle system, and
$F(0)=\big(1,F_{1}(0,x_{1}),\ldots,F_{s}(0,x_{1},\ldots,x_{s}),\ldots\big)$
is a sequence  of initial $s$-particle distribution functions
$F_{s}(0,x_{1},\ldots,x_{s})$
 def\/ined on the phase space
$\mathbb{R}^s\times(\mathbb{R}^s\setminus W_s)$.

Consider the initial value problem for the BBGKY hierarchy
(\ref{BBGKY}), (\ref{BBGKYt0}) with the initial data $F(0)$ from
the space $L_{\xi,\beta}^\infty$ of  sequences
$f=\big(1,f_1(x_1),\ldots,f_n(\qwe x1n),\ldots\big)$ of bounded
functions $f_n(\qwe x1n), \,f_0\equiv1,\, n\geq0,$ that are
def\/ined on the phase space
$\mathbb{R}^n\times(\mathbb{R}^n\setminus W_n)$, are invariant
under permutations of the arguments $x_i,$ $i=1,\ldots,n,$ and
are equal to zero on the set $ W_n$. The norm in this space is
def\/ined by the formula
\begin{gather}
\label{norm_infty}
    {\norm f}=
    \sup_{n\geq0}
    \xi^{-n}\,
    \sup_{\qwe x1n}
    |f_n(\qwe x1n)|\,
    \exp\left\{\beta\sum\limits_{i=1}^n\frac{p_i^2}{2}\right\},
\end{gather}
where $\xi$, $\beta$ are positive integers. Note that the
sequences of $n$ particle equilibrium distribution functions of
inf\/inite systems belong to the space $L_{\xi,\beta}^\infty$
\cite{PGM02,Ryu}.

Let $Y\equiv\{x_1,\ldots,x_s\}$, $X\equiv\{Y,\qwe x{s+1}{s+n}\},$
namely, $X\setminus Y=\{\qwe x{s+1}{s+n}\},$ and let the symbol
$|X|=|Y|+|X\setminus Y|=s+n$ denote the number of the elements of
the set $X.$ By the symbol $X_Y$ we denote the set $X$ with the
subset $Y$ treated as a single element similar to
$x_{s+1},\ldots,x_{s+n}$. Let $L_0^1$ be the subspace of f\/inite
sequences of continuously dif\/ferentiable functions with compact
supports of space $L^1$ of sequences of integrable functions. For
$F(0)$ from $L_0^1$, and hence for all $F(0)\in L_0^1\cap
L_{\xi,\beta}^\infty$ it was proved in \cite{GRS04,GR02} that the
solution
$F(t)=\big(1,F_{1}(t,x_{1}),\ldots,F_{s}(t,x_{1},\ldots,x_{s}),\ldots\big)$
of the initial value problem   (\ref{BBGKY}), (\ref{BBGKYt0}) is
determined by the series expansion
\begin{gather}
       F_{|Y|}(t,Y)
      =\sum\limits_{n=0}^{\infty}
       \frac{1}{n!}
       \int\limits_{\RR^n\times(\mathbb{R}^n\setminus W_n)}
       d(\bs XY){}\!
       \sum\limits_{\texttt{P}:\, X_{Y}=\bigcup\limits_lX_l}\!
       (-1)^{|\texttt{P}|-1}(|\texttt{P}| -1)!\,
       \nonumber\\
\phantom{F_{|Y|}(t,Y)=}{}\times
       \prod_{X_l\subset\texttt{P}}
       S_{\abs {X_l}}(-t,X_l)\,
       F_{|X|}(0,X),
       \qquad
       |\bs XY|\geq0, \label{FgA}
\end{gather}
where $\sum\limits_\texttt{P}$ is the sum over all  possible
partitions $\texttt{P}$ of the set $ X_Y$ into $|\texttt{P}|$
 mutually disjoint nonempty subsets
$X_l\subset X_Y, $ $X_k\cap X_l=\varnothing,$ $k\neq l,$ such that the
entire set $ Y $ is contained in one of the subsets $ X_l$.

On the set of  sequences $f\in L_0^1\cap L_{\xi,\beta}^\infty$,
the evolution operator $S_{|X_l|}(-t,X_l)$ from
expansion~(\ref{FgA}) is given by the formula
\begin{gather}
    \big(S(-t)f\big)_{|X_l|}(X_l)=
    \bigl( S_{|X_l|}\!(-t)f_{|X_l|}\bigr)\!(X_l)
    \equiv
    S_{|X_l|}(-t,X_l)\,
    f_{|X_l|}(X_l)\nonumber\\
\qquad {} =
    \begin{cases}
         f_{|X_l|}\big(\texttt{X}_{1}(-t,X_l),\ldots,\texttt{X}_{|X_l|}(-t,X_l)\big),
         &\text{if
         $x\!\in\!\left(\RR^{|X_l|}\!
         \times
         \big(\RR^{|X_l|}\setminus{W}_{|X_l|}\big)\right)\setminus{\mathcal{M}}_{|X_l|}^0 $}, \\
         0,
         &\text{if
         $x\!\in\!\RR^{|X_l|}\!\times{W}_{|X_l|} $},
    \end{cases}\label{S}
\end{gather}
where $\texttt{X}_{1}(-t,X_l),$ $i=1,\ldots,|X_l|,$ is the
solution of the initial value problem for the Hamilton equations
of the system of $|X_l|$ particles with initial data
$\texttt{X}_{i}(0,X_l)=x_i$ ($S_{|X_l|}(0)=I$ is the identity
operator). Under conditions (\ref{Phi}) on the potential $\Phi$,
the evolution operator (\ref{S}) exists for
$t\in(-\infty,+\infty)$;
 its properties are described in~\cite{PGM02}.

 In the $n$th term of expansion (\ref{FgA}), the form of the integrands is
 constructed by using the cumulant of order $1+n$  for the evolution operators
 (\ref{S}):
\begin{gather}
\label{gA}
    \sum\limits_{\texttt{P}:\, X_{Y}=\bigcup\limits_lX_l}\!
    (-1)^{|\texttt{P}|-1}(|\texttt{P}|-1)!\,
    \prod_{X_l\subset\texttt{P}}S_{\abs {X_l}}(-t,X_l)\equiv
    \gA_{1+\abs{\bs XY}}(t,X_Y),\qquad
    |\bs XY|\geq0.
\end{gather}
Here, the  notation from formula (\ref{FgA}) is  used. Note that
the order of the  cumulant $\gA_{1+|\bs XY|}(t)$ is determined by
the number of  elements of the set $ X_{Y}$ (in this case, by
$1+|X\setminus Y|$ elements).

The simplest examples ($1+\abs{\bs XY}=1,2,3$) of the evolution
operator $ \mathfrak{A}_{1+n}(t) $ (\ref{gA}) have the form
\cite{GR02,Co62a,GrPi63} \setcounter{equation}{7}
\begin{subequations}
\begin{gather}
\label{gAa} \gA_{1}(t,Y)=
    S_{s}(-t,Y),\\
\label{gAb} \gA_{2}(t,Y,x_{s+1})=
    S_{s+1}(-t,Y,x_{s+1})-
    S_s(-t,Y)\,
    S_1(-t,x_{s+1}), \\
\gA_{3}(t,Y,x_{s+1}, x_{s+2})=
    S_{s+2}(-t,Y,x_{s+1}, x_{s+2})
   -S_{s+1}(-t,Y,x_{s+1}) \,
    S_{1}(-t,x_{s+2})                   \nonumber \\
\qquad{} -S_{s+1}(-t,Y,x_{s+2}) \,
    S_{1}(-t,x_{s+1})
   -S_{s}(-t,Y) \,
    S_{2}(-t,x_{s+1},x_{s+2})+           \nonumber\\
\qquad{} +2!S_{s}(-t,Y) \,
    S_{1}(-t,x_{s+1}) \,
    S_{1}(-t,x_{s+2}).
\end{gather}
\end{subequations}
Thus, the cumulant representation of the solution (\ref{FgA})  of
the initial value problem for the BBGKY hierarchy (\ref{BBGKY}),
(\ref{BBGKYt0}) is determined by
 the cumulants
$\gA_{1+|\bs XY|}(t)$ (\ref{gA}) for the evolution operators
$S_{|X_l|}(-t,X_l)$~(\ref{S}).

\section{Regularization of solution}

For $F(0)\in L_{\xi,\beta}^\infty$ every term $n\equiv |X\setminus
Y|$ of expansion (\ref{FgA}) %
contains divergent integrals with respect to the conf\/iguration
variables.  Let us show that the above-stated cumulant nature of
the solution expansions (\ref{FgA}) for the initial value problem
of the BBGKY hierarchy (\ref{BBGKY}), (\ref{BBGKYt0}) guarantees
the compensation of the divergent integrals, i.e., the cumulants
are determined terms of expansion~(\ref{FgA}) as the sum of
summands with divergent integrals that compensate one another. In
order to prove this fact, let us rearrange the terms of
expansion~(\ref{FgA}) so that they are represented by the simplest
mutually compensating groups of summands.  Such procedure will be
called a~regularization of the solution~(\ref{FgA}).  In this
case, the regularization will be based on expressing cumulants of
higher order in terms of the f\/irst and second order cumulants.
For f\/ixed initial data the second-order cumulants will be
determined by expressions which compensate each other over a
certain bounded domain.

The next lemma shows that, in the general case, the cumulant of
the $(1+n)$th order $\gA_{1+n}(t),$ $n\geq1,$ is expressed via the
f\/irst and second order cumulants.

\begin{lemma}
\label{L:reg}
   The equality
    \begin{gather}
        \gA_{1+\abs{\bs XY}}(t,X_Y)
       =\sum\limits_{\substack{Z\subset\bs XY\\Z\neq\varnothing}}
        \gA_{2}(t,Y,Z)\nonumber\\
\phantom{\gA_{1+\abs{\bs XY}}(t,X_Y) =}{}\times
        \sum\limits_{\texttt{\tt P}:\bs X{(Y\cup Z)}=\bigcup\limits_lX_l}
        (-1)^{|\texttt{\tt P}|}(|\texttt{\tt P}|)!\,
        \prod_{\substack{X_l\subset \texttt{\tt P}\\} }
        \gA_{1}(t,X_l),
        \qquad
        |\bs XY|\geq1,    \label{gA01}
    \end{gather}
is true,    where
    $\sum\limits_{Z}$
    is  the sum over all the nonempty  subsets
    $Z$
    of the set
    $\bs XY,$ $Z\subset\bs XY,$
    and the group of
    $|Z|$
    particles evolves as a single element, and
    $\sum\limits_{\texttt{P}}$
    is  the sum over all possible partitions
    $\texttt{\tt P}$
    of the set
    $ \bs X{(Y\cup Z)}$
    into
    $|\texttt{\tt P}|$
    mutually disjoint nonempty subsets
    $X_l\subset\bs X{(Y\cup Z)},$
    $X_k\cap X_l=\varnothing$, $k\neq l,$
    such that  every cluster of
    $|X_l|$
    particles evolves as a single element.
\end{lemma}

The proof of Lemma \ref{L:reg} is based on the verif\/ication that
(\ref{gA01}) is equal to expression (\ref{gA}) by taking into
account the representation
 of a
second order cumulant in terms  of  the f\/irst order ones.

By using Lemma \ref{L:reg}, represent integrands of every summand
from expansion (\ref{FgA}) in terms of the f\/irst  and second
order cumulants.
 As
a result,  expansion (\ref{FgA}) for the solution of the initial
value problem (\ref{BBGKY}), (\ref{BBGKYt0}) takes the form
\begin{gather}
\label{FgA12}
       F_{|Y|}(t,Y)
      =\gA_{1}(t,Y)
       F_{|Y|}(0,Y)
      +\sum\limits_{n=1}^{\infty}
       \frac{1}{n!}\!
       \int\limits_{\RR^n\times(\mathbb{R}^n\setminus W_n)}
       d(\bs XY){}
      \\
       {}\times
       \sum\limits_{\substack{Z\subset\bs XY\\Z\neq\varnothing}}
       \gA_{2}(t,Y,Z)\!
       \sum\limits_{\texttt{P}:\bs X{(Y\cup Z)}=\bigcup\limits_lX_l}
       (-1)^{|\texttt{P}|}
       |\texttt{P}|!\,
       \prod_{\substack{X_l\subset \texttt{P}\\} }
       \gA_{1}(t,X_l)\,
       F_{|X|}(0,X),
       \quad
       |\bs XY|\geq1. \nonumber
\end{gather}

For the initial data $F(0)\in L^1\cap L_{\xi,\beta}^\infty$, the
equality
\begin{gather}
       \int\limits_{\RR^n\times(\mathbb{R}^n\setminus W_n)}\!\!\!
       d(\bs XY){}\!
       \sum\limits_{\substack{Z\subset\bs XY\\Z\neq\varnothing}}
       \gA_{2}(t,Y,Z)\!\!\!
       \sum\limits_{\texttt{P}:\bs X{(Y\cup Z)}=\bigcup\limits_lX_l}
       (-1)^{|\texttt{P}|}
       |\texttt{P}|!\,
       \prod_{\substack{X_l\subset \texttt{P}\\} }
       \gA_{1}(t,X_l)\,
       F_{|X|}(0,X)\nonumber\\
       =
       \int\limits_{\RR^n\times(\mathbb{R}^n\setminus W_n)}
       d(\bs XY){}\!
       \sum\limits_{\substack{Z\subset\bs XY\\Z\neq\varnothing}}
       (-1)^{|\bs X{(Y\cup Z)}|}
       \gA_{2}(t,Y,Z)\,
       F_{|X|}(0,X),
       \qquad
       |\bs XY|\geq1,\label{intgA12=}
\end{gather}
is true. Here, we have used
 the  Liouville theorem \cite{PGM02} and taken into account the relation
\begin{gather}
\label{s(m,k)}
    \sum\limits_{k=1}^m
    (-1)^k
    k!\,
    s(m,k)=
    (-1)^m,
    \qquad
    m\geq1,
\end{gather}
where $s(m,k)$ is  the Stirling number  of the second kind
def\/ined as the number of all distinct  partitions of a set
containing  $m$ elements into $k$ subsets.

Thus, by virtue of equality (\ref{intgA12=}), the expansion
(\ref{FgA12}) for the solution of the initial value problem for
the BBGKY hierarchy (\ref{BBGKY}), (\ref{BBGKYt0}) takes the form
\begin{gather}
\label{FgA2}
       F_{|Y|}(t,Y)
      =\gA_{1}(t,Y)
       F_{|Y|}(0,Y)\\
    {}  +\sum\limits_{n=1}^{\infty}
       \frac{1}{n!}\!\!
       \int\limits_{\RR^n\times(\mathbb{R}^n\setminus W_n)}\!\!\!
       d(\bs XY){}\!
       \sum\limits_{\substack{Z\subset\bs XY\\Z\neq\varnothing}}
       (-1)^{|\bs X{(Y\cup Z)}|}
       \gA_{2}(t,Y,Z)\,
       F_{|X|}(0,X),
       \qquad
       |\bs XY|\geq1,\nonumber
\end{gather}
where the notation from (\ref{gA01}) is used.

Representation (\ref{FgA2}) 
will be called a regularized cumulant representation of the
solution of the initial value  problem for the BBGKY hierarchy
(\ref{BBGKY}), (\ref{BBGKYt0}).

\section{Existence theorem}

Taking into account the invariance of the Gibbs distributions (the
Maxwellian distribution can be extended to become a Gibbs
distribution) with respect to the action of evolution operators
$S\left( - t \right)$~(\ref{S}) and using the
condition~(\ref{Phii}) and the relation (\ref{gA01}) we estimate
the integrands in expansion (\ref{FgA2}).

\begin{lemma}
\label{L:gA2F<=}
    If
    $F(0)\in L_{\xi,\beta}^\infty$
    then the  inequality
    \begin{gather}
           \left|
       \sum\limits_{\substack{Z\subset\bs XY\\Z\neq\varnothing}}
       (-1)^{|\bs X{(Y\cup Z)}|}
       \gA_{2}(t,Y,Z)\,
       F_{|X|}(0,X)
       \right|
       \nonumber\\
\qquad{}
       \leq
       2\norm{F(0)}
       (\xi\,\mathrm{e}^{2\beta b})^{s}\,
       (\xi\,\mathrm{e}^{2\beta b})^{n}
       \exp\left\{-\beta\sum\limits_{i=1}^{s+n}\frac{p_i^2}{2}\right\}\label{gA2F<=}
    \end{gather}
    holds, where the notation from formulae \eqref{Phii},
\eqref{norm_infty} and \eqref{gA01} is used.
\end{lemma}

As a consequence of Lemma~\ref{L:gA2F<=}, the following existence
theorem is true.

\begin{theorem*}
    If
    $F(0)\in L_{\xi,\beta}^\infty$
    is a sequence of nonnegative functions, then for
    $
    \xi<\frac{\mathrm{e}^{-2\beta b-1}}{2\tilde{C_1}}
    \sqrt{\frac{\beta^{\prime\prime}}{2\pi}}$
    and
    $t\in[0,t_0)$,
    where
    $
    t_0=
    \frac{1}{\tilde{C_2}}
    \Big(
    \frac
    {\mathrm{e}^{-2\beta b-1}}
    {2\xi}
    \sqrt{\frac{\beta^{\prime\prime}}{2\pi}}-
    \tilde{C_1}
    \Big),$
    $\tilde{C_1}=\max(2R,1)$,
    $\tilde{C_2}=\max\big(2(4b+1),\frac{2}{\beta^{\prime}}\big),$
    $\beta=\beta^\prime+\beta^{\prime\prime}$,
    $b\equiv \underset{q\in[\sigma,R]}{\sup}\big|\Phi(q)\big|\big(\big[\frac{R}{\sigma}\big]\big)$
and    $\big[\frac{R}{\sigma}\big]$
    is the integer part  of the number
    $\frac{R}{\sigma}$,
    there exists a unique weak solution
    of the initial value problem for the BBGKY hierarchy
    \eqref{BBGKY}, \eqref{BBGKYt0},
    namely, the sequence
    $F(t)\in L_{\xi,\beta}^\infty$
    of nonnegative functions
    $F_s(t)$
    determined by  expansion \eqref{FgA2}.
    \end{theorem*}

\begin{proof}
        Let particles interact via a short-range pair potential that satisf\/ies
        conditions (\ref{Phi}).
        We assume that, at the initial instant, the conf\/iguration coordinates
        $q_i$, $i=1,2,\dots,s$, of the particles constituting the
        cluster $Y$ take values in a compact set of those $|Y|$ intervals $l_{i}$
        with length $|l_{Y}|$
        that
        $
            q_i\in l_{i}.
        $
        Then if during the time interval
        $[0,t)$ none of the particles of any cluster
        $Z\subset\bs XY$
        interacts  with the particles of cluster $Y$,
        then the  operator equality holds
\[
S_{|Y\cup Z|}(-t,Y, Z)=S_{|Y|}(-t,Y)S_{|Z|}(-t,Z),
\]
        and, as a result,  we have
\[
\gA_{2}(t,Y,Z)F_{|X|}(0,X)=0.
\]
        Therefore, in this case, the integrands in the $n$th term of expansion
        (\ref{FgA2}) are equal to zero.  Since $Z\subset\bs XY$, it follows
        that in the expansion (\ref{FgA2}) the domain $\mathbb{R}^n\setminus W_n$
        of integration with respect to the conf\/iguration
        variables is determined by $n$ bounded intervals where the particles
        of the cluster $Y$ during the time interval $[0,t)$ interact with
        particles of the cluster $Z$ that contains the maximal number of
        particles, namely, with the particles of the cluster $\bs XY$. Thus, the
        domain $\mathbb{R}^n\setminus W_n$ of integration estimates
        the following f\/inite value
        \begin{gather}
        \label{V<=}
            V(t)
            \leq
            \left(
            C+C_0t+(C_1+C_2t)n+t\sum_{i=s+1}^{s+n}p_i^2
            \right)^{n},
        \end{gather}
        where
        $
        C\equiv |l_{Y}| +2sR$,
        $C_0\equiv 2s(4b+1)+\sum\limits_{i=1}^{s}p_i^2,$
        $C_1\equiv 2R$,
        and  $C_2\equiv2(4b+1).
        $

        In view of estimates (\ref{gA2F<=}) and (\ref{V<=}), we obtain
        \begin{gather}
            \label{majF}
              |F_{|Y|}(t,Y)|
              \leq
              2\norm{F(0)}\,
              (\xi\mathrm{e}^{2\beta b})^s
              \exp
              \left\{
                -\beta\sum_{i=1}^{s}
                \frac{p_i^2}{2}
              \right\}
              \sum_{n=0}^\infty
              \frac{1}{n!}
              (\xi\mathrm{e}^{2\beta b})^n
              \\
              \qquad{}\times
              \int_{\RR^n}
              d p_{s+1}
              \cdots
              d p_{s+n}
              \exp
              \left\{
                -\beta
                \sum_{i=s+1}^{s+n}
                \frac{p_i^2}{2}
              \right\}
              \left(C+C_0t+(C_1+C_2t)n+t\sum_{i=s+1}^{s+n}p_i^2\right)^{n}.\nonumber
            \end{gather}

         Taking in account the relation
            \begin{gather}
                            \left((C+C_0t)+(C_1+C_2t)n+t\sum_{i=s+1}^{s+n}p_i^2\right)^{n}\nonumber\\
              \qquad{} =\sum_{k=0}^n\frac{n!}{k!}(C+C_0t)^k
                \sum_{r=0}^{n-k}
                \frac{1}{r!}
                \big((C_1+C_2t)n\big)^r
                \frac{1}{(n-k-r)!}
                t^{n-k-r}
                \left(\sum\limits_{i=s+1}^{s+n} p_i^2\right)^{n-k-r}\label{VtoN}
            \end{gather}
        and the  inequality
        \begin{gather*}
            \label{pexp<=}
                \left(\sum\limits_{i=s+1}^{s+n} p_i^2\right)^{n-k-r}
                \exp\left\{-\beta^\prime\sum_{i=s+1}^{s+n}\frac{p_i^2}{2}\right\}
                \leq
                (n-k-r)!
                \left(\frac{2}{\beta^\prime}\right)^{n-k-r},
            \end{gather*}
            we compute the integrals with respect to the momentum
            variables in the right-hand side of (\ref{majF}):
            \begin{gather}
            \label{Int_p}
              \int\limits_{\RR^n}
              d p_{s+1} \cdots dp_{s+n}
              \exp\bigg\{-\beta^{\prime\prime}\sum_{i=s+1}^{s+n}\frac{p_i^2}{2}\bigg\}=
              \left(\frac{2\pi}{\beta^{\prime\prime}}\right)^{\frac{n}{2}},
              \qquad
            \beta=\beta^\prime+\beta^{\prime\prime}.
            \end{gather}

        Then estimate (\ref{majF})  takes the form
         \begin{gather}
                         |F_{|Y|}(t,Y)|
              \leq
              2\norm{F(0)}\,
              (\xi\mathrm{e}^{2\beta b})^s
              \exp
              \left\{
                -\beta\sum_{i=1}^{s}
                \frac{p_i^2}{2}
              \right\}
              \nonumber\\
              \qquad{}\times
              \sum_{n=0}^\infty
              (2\xi\mathrm{e}^{2\beta b})^n
              \left(\frac{2\pi}{\beta^{\prime\prime}}\right)^{\frac{n}{2}}
              \sum_{k=0}^n\frac{(C+C_0t)^k}{k!}
                \sum_{r=0}^{n-k}
                \frac{n^r}{r!}
                (C_1+C_2t)^r
                \frac{t}{(n-k-r)!}.
                \tag{\ref{majF}$'$}\label{majF'}
        \end{gather}
        Let us  put
$ \tilde{C_1}=\max(C_1,1)$ and
        $\tilde{C_2}=\max\Big(C_2,\frac{2}{\beta^{\prime}}\Big)  $
and continue estimate (\ref{majF'}).
        For arbitrary $t\geq0$, the inequalities  $
        \tilde{C_1}+\tilde{C_2t}\geq1 $ and $(\tilde{C_1}+\tilde{C_2}t)
        \frac{2t}{\beta^{\prime}}\geq1 $ are true, and, therefore,
\[
        (C_1+C_2t)^r
        \left(\frac{2t}{\beta^{\prime}}\right)^{n-k-r}
        \leq
        (\tilde{C_1}+\tilde{C_2}t)^n.
\]
        By using the inequalities
        $
        \sum\limits_{r=0}^{n-k}
        \frac{n^r}{r!}
        \leq
        \mathrm{e}^n
$ and
 $       \sum\limits_{k=0}^n
        \frac{(C+C_0t)^k}{k!}
        \leq
        \mathrm{e}^{(C+C_0t)}
        $
        and taking estimate (\ref{majF'}) into account, we obtain
        \begin{gather}
                          \abs{F_{|Y|}(t,Y)}
              \leq
              2\norm{F(0)}
              (\xi\mathrm{e}^{2\beta b})^s
              \exp
              \left\{
                -\beta\sum_{i=1}^{s}
                \frac{p_i^2}{2}
              \right\}\nonumber\\
 \phantom{\abs{F_{|Y|}(t,Y)}\leq }{}\times
              \mathrm{e}^{(C+C_0t)}
              \sum_{n=0}^\infty
              \left(
              2\xi\mathrm{e}^{2\beta b+1}
              \sqrt{\frac{\beta^{\prime\prime}}{2\pi}}
              \right)^n
              (\tilde{C_1}+\tilde{C_2}t)^n.\label{majjF}
        \end{gather}

        Thus, if
        $
        \xi<\frac{\mathrm{e}^{-2\beta b-1}}{2\tilde{C_1}}
        \sqrt{\frac{\beta^{\prime\prime}}{2\pi}}
        $
        then series (\ref{majjF}) converges for
        $
        0\leq t<
        t_0\equiv
        \frac{1}{\tilde{C_2}}
        \Big(
        \frac
        {\mathrm{e}^{-2\beta b-1}}
        {2\xi}
        \sqrt{\frac{\beta^{\prime\prime}}{2\pi}}-
        \tilde{C_1}
        \Big).
        $
        We have thus shown that,
under the conditions
        of the
theorem,
  series (\ref{majF}) converges.

Finally,  using  the well-known theorems of functional analysis
\cite{RS72}  and arguing
  similarly to~\mbox{\cite{PGM02, CGP97}},
we show that the sequence $F(t)$ is the unique weak solution of
the
  initial value problem for the BBGKY hierarchy
  (\ref{BBGKY}), (\ref{BBGKYt0}).
\end{proof}

\section{Conclusion}
The theory of the BBGKY hierarchy is developing now since the area
of its application grows~\cite{Kan03,Tar03}. At present while
solving the initial value problem of the BBGKY hierarchy, many
mathematical problems arise \cite{CIP94,Pe79,IP86,IP87}. \ In this
paper one of such problems is considered for inf\/inite particle
system employing the method of the interaction region \cite{Pe79}.
Taking into account the cumulant representation  \cite{GRS04,GR02}
we construct a new regularized representation of the solution of
the BBGKY hierarchy for a one-dimensional inf\/inite system of
hard spheres interacting via a short-range potential. For the
initial data from the space of sequences of functions which are
bounded in conf\/iguration variables and exponentially decreasing
in momentum variables, existence of the cumulant representation
allows to properly regularize such expansion for the solution,
determining by second order cumulant as the sum of summands with
divergent integrals that compensate one another. A
multidimensional case and a case of a more general interaction
potential  will be investigated in further contribution.

\subsection*{Acknowledgement}

 The author is pleased to thank Professor Victor~Gerasimenko for many
useful discussions and is grateful to the referees for helpful
comments and references. This work was partially supported by the
National Academy of   Sciences of Ukraine through grant
No~0105U005666 for young
    scientists and partially supported by INTAS, grant No 00-0015.

\LastPageEnding

\newpage

\begin{thebibliography}{99}
\footnotesize

\bibitem{PGM02}
        Petrina D.Ya., Gerasimenko V.I., Malyshev P.V.,
        Mathematical foundations of classical statistical mechanics.
        Continuous systems, 2nd ed.,  London~-- New York, Taylor \& Francis Inc., 2002.

\bibitem{CGP97}
        Cercignani C., Gerasimenko V.I., Petrina D.Ya.,
        Many-particle dynamics and kinetic equations,
        Kluwer Acad. Publ., 1997.

\bibitem{CIP94}
        Cercignani C., Illner R., Pulvirenti M.,
        The mathematical theory of dilute gases,
        {\it Applied Mathematical Sciences}, Vol.~106,
        New York, Springer, 1994.

\bibitem{Sp91}
        Spohn H.,
        Large scale dynamics of interacting particles, Springer, 1991.

\bibitem{Pe79}
        Petrina D.Ya.,
        Mathematical description of the evolution of inf\/inite
        systems of classical statistical physics. I. Locally perturbed one-dimensional systems,
        {\it Teoret. Mat. Fiz.}, V.38, 1979, 230--262 (in Russian).

\bibitem{PG83}
        Petrina D.Ya., Gerasimenko V.I.,
        Mathematical description of the evolution of the state of inf\/inite systems of classical statistical mechanics,
        {\it Uspekhi Mat. Nauk}, 1983, V.38, 3--58 (in Russian).

\bibitem{GR03}
        Gerasimenko V.I., Ryabukha T.V.,
        Dual nonequilibrium cluster expansions,
        {\it Dopov. Nats. Akad. Nauk Ukr. Mat. Prirodozn. Tekh. Nauki}, 2003,  N~3, 16--22 (in Ukrainian).

\bibitem{GRS04}
        Gerasimenko V.I., Ryabukha T.V., Stashenko M.O.,
        On the BBGKY hierarchy solutions for many-particle systems with dif\/ferent symmetry properties,
        in Proceedings of Fifth International Conference ``Symmet\-ry in
Nonlinear Mathematical Physics'' (June 23--29, 2003, Kyiv),
Editors A.G. Nikitin, V.M.~Boyko, R.O.~Popovych and
I.A.~Yehorchenko, {\it Proceedings of Institute of Mathematics},
Kyiv, 2004, V.50, Part~3, 1308--1313.

\bibitem{GR02}
        Gerasimenko V.I., Ryabukha T.V.,
        Cumulant representation of solutions of the Bogolyubov chains of equations,
        {\it Ukrain. Mat. Zh.}, 2002, V.54, 1313--1328
        (English transl.: {\it Ukrainian Math. J.},
         2002, V.54, 1583--1601).

\bibitem{Ryu}
        Ruelle D.,
        Statistical mechanics. Rigorous results,
        New York~-- Amsterdam, W.A.~Benjiamin Inc., 1969.

\bibitem{Co62a}
        Cohen E.G.D.,
        Cluster expansions and the hierarchy. I. Non-equilibrium distribution functions,
        {\it Physica}, 1962, V.28, 1045--1059.

\bibitem{GrPi63}
        Green H.S., Piccirelli R.A.,
        Basis of the functional assumption in the theory of the Boltzmann equation,
        {\it Phys. Rev. (2)}, 1963, V.132, 1388--1410.

\bibitem{RS72}
        Reed M., Simon B.,
        Methods of modern mathematical physics. Vol.~1: Functional analysis,
        New York~-- London, Academic Press, 1972.

\bibitem{Kan03}
        Kaniadakis G.,
        BBGKY hierarchy underlying many-particle quantum mechanics,
        {\it Phys. Lett. A}, 2003, V.310, 377--382,  quant-ph/0303159.

\bibitem{Tar03}
        Tarasov V.E.,
        Fractional systems and fractional Bogoliubov hierarchy equations,
        {\it Phys. Rev.~E}, 2005, V.71,  011102, 12 pages, cond-mat/0505720.

\bibitem{IP86}
        Illner R., Pulvirenti M.,
        Global validity of the Boltzmann equation for a two-dimensional rare gas in vacuum,
        {\it Comm. Math. Phys.}, 1986, V.105, 189--203.

\bibitem{IP87}
        Illner R., Pulvirenti M.,
        A derivation of the BBGKY-hierarchy for hard sphere particle systems,
        {\it  Transport Theory Statist. Phys.}, 1987, V.16, 997--1012.
\end{thebibliography}
\end{document}